\documentclass[12pt]{article}
\usepackage{graphicx}
\begin{document}

\title{Diffusion laws, path information and action principle}

\author{Qiuping A. Wang\\
{\it Institut Sup\'erieur des Mat\'eriaux et M\'ecaniques Avanc\'es}, \\
{\it 44, Avenue F.A. Bartholdi, 72000 Le Mans, France}}

\date{}

\maketitle

\begin{abstract}
This is an attempt to address diffusion phenomena from the point of view of
information theory. We imagine a regular hamiltonian system under the random
perturbation of thermal (molecular) noise and chaotic instability. The irregularity
of the random process produced in this way is taken into account via the dynamic
uncertainty measured by a path information associated with different transition
paths between two points in phase space. According to the result of our previous
work, this dynamic system maximizes this uncertainty in order to follow the action
principle of mechanics. In this work, this methodology is applied to particle
diffusion in external potential field. By using the exponential probability
distribution of action (least action distribution) yielded by maximum path
information, a derivation of Fokker-Planck equation, Fick's laws and Ohm's law for
normal diffusion is given without additional assumptions about the nature of the
process. This result suggests that, for irregular dynamics, the method of maximum
path information, instead of the least action principle for regular dynamics, should
be used in order to obtain the correct occurring probability of different paths of
transport. Nevertheless, the action principle is present in this formalism of
stochastic mechanics because the average action has a stationary associated with the
dynamic uncertainty. The limits of validity of this work is discussed.

\end{abstract}


\section{Introduction}
Diffusion is a mechanism by which components of a mixture are transported around the
mixture by means of random molecular motion. Over 200 years ago, Berthalot
postulated\cite{Berthalot} that the flow of mass by diffusion across a plane was
proportional to the concentration gradient of the diffusant across that plane. About
50 years later, Fick introduced\cite{Fick} two differential equations that quantified
the above statement for the case of transport through thin membranes. Fick's First Law
states that the flux $J$ of a component of concentration $n$ across a membrane is
proportional to the concentration gradient in the membrane:
\begin{eqnarray}                                            \label{1}
J(x)=-D\frac{\partial n(x)}{\partial x}
\end{eqnarray}
where $x$ is the position variable (for 3 dimensional space,
$\frac{\partial}{\partial x}$ is replaced by the gradient $\nabla$). Fick's Second
Law states that the rate of time change of concentration of diffusant at a point is
proportional to the rate of spacial change of concentration gradient at that point
within the mixture
\begin{eqnarray}                                            \label{2}
\frac{\partial n}{\partial t} = \frac{\partial}{\partial x}[D\frac{\partial
n(x)}{\partial x}].
\end{eqnarray}
If $D$ is constant everywhere in the mixture, the above equation becomes
$\frac{\partial n}{\partial t} = D\frac{\partial^2 n(x)}{\partial x^2}$.

Above normal diffusion laws are very precisely tested in experiments in most of
solids, liquids and gases and are widely studied in nonequilibrium thermostatistics
together with the Fokker-Planck equation of diffusion probability, Fourier law of heat
conduction and Ohm's law of electrical charge
conduction\cite{Lebowitz,Bonetto,Bonetto2,Ohm}. Many efforts to derive theoretically
Fick's laws were concentrated on special models of solids in which particles are
transported. Other phenomenological derivations are also possible if one supposes
Brownian motion and Markovian process\cite{Kubo}, or Kolmogorov
conditions\cite{Zaslavsky}. Although diffusion is associated with stochastic process
with certain dynamic uncertainty, theoretical study and derivation of these laws using
information methods, as far as we know, is still an open question.

In this paper, we describe an attempt to investigate irregular dynamics by an
information method. We imagine a hamiltonian systems under random perturbation of
thermal noise and chaotic motion. This perturbation can be either internal or
external. A isolated system containing a large number of particles in random motion is
an example of internal perturbation. A Brownian particle is an example of external
perturbation due to random molecular motion around the particle. In these cases, the
energy of the system and other mechanical quantities can fluctuate. The geodesics of
regular dynamics will be deformed in such a stochastic way that following exactly the
evolution of each mechanical quantity (e.g., action) is practically inconceivable.
Hence in this approach, we use the action of the unperturbed hamiltonian system. The
stochastic effect of random perturbation is taken into account through the dynamic
uncertainty or information. Based on maximum information principle in connection with
the unperturbed action, we developed a formalism of probabilistic dynamics within
which the diffusion laws are a natural result of the differential equations of the
probability distribution of action. The information addressed in this work is a
quantification of the uncertainty of irregular dynamic process. Here we consider the
following two dynamic uncertainties:

\begin{enumerate}

\item Between any two fixed cells $a$ and $b$ in phase space, there may be different
possible paths (labelled by $k$=1,2,...w) each having a probability $p_k(b|a)$ to be
followed by the system. This is the uncertainty considered by Feynman in his
formulation of quantum mechanics\cite{Feynman}. Here we introduce it for randomly
perturbed dynamic systems.

\item There are different possible paths leaving the cell $a$ and leading to different
final cells $b$ in a final phase volume $B$, each having a probability $p_k(x|a)$ to
be followed by the system, where $x$ is the position of arbitrary $b$. This
uncertainty is the basic consideration for the definition of Kolmogorov-Sinai
entropy\cite{Dorfman} used to describe chaotic systems.

\end{enumerate}

These dynamic uncertainties were proved\cite{Wang04x,Wang04xx} to take their maximum
when the most probable paths are just the paths of least action. In what follows, we
briefly present the method of maximum uncertainty or path information for irregular
dynamic systems and the concomitant ``least action distributions'' of transition
probability. Then we focus on the particles diffusing in external field with potential
energy $U(x)$. The Hamiltonian of the particles is given by $H=E+U$ where $E$ is the
kinetic energy. The above mentioned diffusion laws will be derived directly from the
so called least action distribution.

\section{Maximum path information}
We suppose that the uncertainty concerning the choice of paths by the systems
between two $fixed$ cells is measured with the following path information
\begin{eqnarray}                                            \label{c1x}
H_{ab}=-\sum_{k=1}^wp_k(b|a)\ln p_k(b|a).
\end{eqnarray}
where $p_k(b|a)$ is the transition probability between the cell $a$ in the initial
phase volume $A$ and a cell $b$ in the final phase volume $B$ via a path $k$
($k=1,2,...w$). We have the following normalization
\begin{eqnarray}                                            \label{c1xx}
\sum_{k=1}^{w}p_k(b|a)=1.
\end{eqnarray}
It is also supposed that each path is characterized by its action $A_{ab}(k)$
defined in the same way as in regular dynamics without random forces by
\begin{eqnarray}                                            \label{c5}
A_{ab}(k)=\int_{t_{ab}(k)}L_k(t)dt
\end{eqnarray}
where $L_{k}(t)=E-U$ is the Lagrangian of the system at time $t$ along the path $k$.
The average action is given by
\begin{eqnarray}                                            \label{c1xxx}
A_{ab}=\sum_{k=1}^wp_k(b|a)A_{ab}(k).
\end{eqnarray}

It is known that, for a regular process, the trajectories of the system should have a
stationary action ($\delta A_{ab}(k)=0$). For irregular dynamic process, this least
action principle does not apply since there are many possible paths with different
actions, not only the minimum action. For this kind of process, we suppose the stable
probability distribution corresponds to a stationary uncertainty or path information
$H_{ab}$ under the constraint associated with action. This means the following
operation:
\begin{eqnarray}                                            \label{xc1x}
\delta [H_{ab}+\alpha\sum_{k=1}^{w}p_k(b|a)-\eta\sum_{k=1}^wp_k(b|a)A_{ab}(k)]=0
\end{eqnarray}
leading to
\begin{eqnarray}                                            \label{c6x}
p_k(b|a)=\frac{1}{Z}\exp[-\eta A_{ab}(k)],
\end{eqnarray}
where the partition function
\begin{eqnarray}                                            \label{cx6x}
Z=\sum_{k}\exp[-\eta A_{ab}(k)].
\end{eqnarray}
The physical meaning of $\eta$ will be discussed in the following section.

It is proved that\cite{Wang04x} the distribution Eq.(\ref{c6x}) is stable with
respect to the fluctuation of action. It is also proved that Eq.(\ref{c6x}) is a
least (stationary) action distribution, i.e., the most probable paths are just the
paths of least action. We can write $\delta p_k(b|a)=-\eta p_k(b|a)\delta
A_{ab}(k)=0$, which means $\delta A_{ab}(k)=0$ leading to Euler-Lagrange
equation\cite{Arnold}
\begin{eqnarray}                                            \label{wc6}
\frac{\partial}{\partial t}\frac{\partial L_{k}(t)}{\partial \dot{x}}-\frac{\partial
L_{k}(t)}{\partial x}=0
\end{eqnarray}
and to Hamiltonian equations via the Legendre transformation
$H=P\dot{x}-L_k(t)$\cite{Arnold}:
\begin{eqnarray}                                            \label{w6x}
\dot{x}=\frac{\partial H}{\partial P} \;\;and \;\; \dot{P}=-\frac{\partial
H}{\partial x},
\end{eqnarray}
where $P=m\dot{x}$ is the momenta of the system. These equations are satisfied by
the most probable paths but not the other paths. In general, the paths have neither
$\delta A_{ab}(k)=0$ nor $\delta A_{ab}=0$, although the average action $A_{ab}$ is
convex (concave) for $\eta>0$ ($\eta<0$)\cite{Wang04x}. Nevertheless, the average
action does have a stationary associated with the stationary of the information
$H_{ab}$ in Eq.(\ref{xc1x}), i.e.,
\begin{eqnarray}                                            \label{ws6x}
-\eta\delta A_{ab}+\delta H_{ab}=0.
\end{eqnarray}
Here we used $\sum_{k=1}^{w}\delta p_k(b|a)=0$. Eq.(\ref{ws6x}) implies that,
although the least action principle of classical mechanics cannot apply when the
dynamics is perturbed by random and instable noise, the maximum path information
introduced above underlies in fact the same physics in which the action principle is
present as a average effect in association with the stationary dynamic uncertainty.
Eq.(\ref{ws6x}) will be used below in order to derive the averaged Euler-lagrange
equation for irregular dynamics.

\section{A calculation of transition probability of diffusion}
In what follows, the action is calculated and analyzed with the Euler method. Let us
look at a particle of mass $m$ diffusing along a given path $k$ from a cell $a$ to a
cell $b$ of its phase space. The path is cut into $N$ infinitesimally small segments
each having a spatial length $\Delta x_i=x_i-x_{i-1}$ with $i=1 ...N$ ($x_0=x_a$ and
$x_N=x_b$). $t=t_i-t_{i-1}$ is the time interval spent by the system on every segment.
The Lagrangian on the segment $i$ is given by
\begin{eqnarray}                                            \label{x9c}
L(x_i,\dot{x}_i,t)=\frac{m(x_i-x_{i-1})^2}{2(t_i-t_{i-1})^2} -\left(\frac{\partial
U}{\partial x}\right)_i\frac{(x_i-x_{i-1})}{2}-U(x_{i-1})
\end{eqnarray}
where the first term on the right hand side is the kinetic energy of the particle, the
second and the third are the average potential energy on the segment $i$. The action
of segment $i$ is given by
\begin{eqnarray}                                            \label{xx9c}
A_i=\frac{m(\Delta x_i)^2}{2t} +F_i\frac{\Delta x_i}{2}t-U(x_{i-1})t,
\end{eqnarray}
where $F_i=-\left(\frac{\partial U}{\partial x}\right)_i$ is the force on the segment
$i$. According to Eq.(\ref{c6x}), the transition probability $p_{i/i-1}$ from
$x_{i-1}$ to $x_i$ on the path $k$ is given by
\begin{eqnarray}                                            \label{xxxc9}
p_{i,k} &=& \frac{1}{Z_i} \exp\left(-\eta\left[\frac{m}{2t}\Delta x_i^2
+F_i\frac{t}{2}\Delta x_i\right]_{k}\right)
\end{eqnarray}
where $Z_i$ is calculated as follows
\begin{eqnarray}                                            \label{xxx9}
Z_i&=&\int_{-\infty}^\infty dx_i\exp\left(-\eta\left[\frac{m}{2t}\Delta x_i^2
+F_i\frac{t}{2}\Delta x_i\right]_{k}\right)\\ \nonumber &=& \exp\left[F_i^2\frac{\eta
t^3}{8m}\right]\sqrt{\frac{2\pi t}{m\eta}}.
\end{eqnarray}
The potential energy of the point $x_{i-1}$ disappears in the expression of $p_{i,k}$
because it does not depend on $x_i$.

At this stage, the physical meaning of $\eta$ can be shown with a general
relationship given by the following calculation of the variance :
\begin{eqnarray}                                            \label{wxxx9}
\overline{\Delta x_i^2}&=&\frac{1}{Z_i}\int_{-\infty}^\infty dx_i\Delta
x_i^2\exp\left(-\eta\left[\frac{m}{2t}\Delta x_i^2 +F_i\frac{t}{2}\Delta
x_i\right]_{k}\right)\\ \nonumber &=& \frac{t}{m\eta}.
\end{eqnarray}
This result can be compared to Brownian motion having $\overline{\Delta
x_i^2}=4Dt$\cite{Kubo}. In this case we have $\eta=\frac{1}{2mD}$. If we still
consider detailed balance\cite{Kubo}, we get $D=\mu k_BT$ where $\mu$ is the mobility
of the diffusing particles, $k_B$ the Boltzmann constant and $T$ the temperature. This
means
\begin{eqnarray}                                            \label{xxx9x}
\eta=\frac{1}{2m\mu k_BT}=\frac{\gamma}{2k_BT}
\end{eqnarray}
where $m\gamma=\frac{1}{\mu}$ is the friction constant of the particles in the
diffusion mixture.

The total action given by
\begin{eqnarray}                                            \label{cc9}
A_{ab}(k)=\sum_{i=1}^NA_i=\sum_{i=1}^N\left[\frac{m(\Delta x_i)^2}{2t}
+F_i\frac{t}{2}\Delta x_i-U(x_{i-1})t\right]_{k}.
\end{eqnarray}
so the transition probability from $a$ to $b$ via the path $k$ is the following:
\begin{eqnarray}                                            \label{c9}
p_k(b|a) &=& \frac{1}{Z} \exp\left(-\eta\sum_{i=1}^N\left[\frac{m(\Delta x_i)^2}{2t}
+F_i\frac{t}{2}\Delta x_i\right]_{k}\right)  \\ \nonumber &=&
p(b|a)^{-1}\prod_{i=1}^N p_{k,i/i-1}
\end{eqnarray}
where
\begin{eqnarray}                                            \label{c9x}
Z&=&\sum_{k=1}^{w}\exp\left(-\eta\sum_{i=1}^N\left[\frac{m(\Delta x_i)^2}{2t}
+F_i\frac{t}{2}\Delta x_i\right]_{k}\right) \\\nonumber &=&\int_{-\infty}^\infty
dx_1dx_2...dx_{N-1}\exp\left(-\eta\sum_{i=1}^N\left[\frac{m(x_i-x_{i-1})^2}{2t}
+F_i\frac{t}{2}(x_i-x_{i-1})\right]_{k}\right)\\\nonumber &=&
\left(\exp\left[F_i^2\frac{\eta t^3}{8m}\right]\sqrt{\frac{2\pi t}{m\eta}}\right)^N
p(b|a)=Z_i^Np(b|a)
\end{eqnarray}
and
\begin{eqnarray}                                            \label{ac9x}
p(b|a)&=& \exp\left[F_i^2\frac{\eta (t_b-t_a)^3}{8m}\right]\sqrt{\frac{m\eta}{2\pi
(t_b-t_a)}}
\\ \nonumber &\times&\exp\left(-\eta\left[\frac{m(x_b-x_a)^2}{2(t_b-t_a)}
+F_i\frac{t_b-t_a}{2}(x_b-x_a)\right]\right).
\end{eqnarray}
In the above calculation, we have fixed the initial point $x_0=x_a$ and the final
point $x_N=x_b$.

Now in order to see the behavior of transition probability with respect to final
point, we have to relax $x_b=x$ and let it vary arbitrarily as other intermediate
points. This implies we take into account the second uncertainty due to chaotic motion
mentioned in the introduction. The corresponding transition probability $p_k(x|a)$
from $a$ to arbitrary $x$ via the path $k$ has been derived with the maximum path
information combined with action\cite{Wang04xx}. Here we only put it as follows:
\begin{eqnarray}                                            \label{acc9x}
p_k(x|a) &=& p(b|a)p_k(b|a) \\
\nonumber &=& \prod_{i=1}^N p_{k,i/i-1},
\end{eqnarray}
which is normalized by
\begin{eqnarray}                                            \label{acc1xx}
\sum_b\sum_{k=1}^{w}p_k(x_b|a)=\int dx_1dx_2...dx_{N-1}dxp_k(x|a)=1.
\end{eqnarray}

\section{A derivation of the Fokker-Planck equation}

The Fokker-Planck equation describes the time evolution of the probability density
function of position and velocity of a particle. This equation can be derived if we
suppose that the diffusion particles follow Brownian motion and Markovian
process\cite{Kubo}, or Kolmogorov conditions\cite{Zaslavsky}. In what follows, it is
shown that this equation can be derived in a quite general way without above
assumptions. It is just the differential equation satisfied by the least action
distributions given by Eq.(\ref{xxxc9}) and Eq.(\ref{acc9x}).

The calculation of the derivatives $\frac{\partial p_{i/i-1}}{\partial t}$,
$\frac{\partial F_ip_{i/i-1}}{\partial x_i}$ and $\frac{\partial^2 p_{i/i-1}}{\partial
x_i^2}$ straightforwardly leads to
\begin{eqnarray}                                            \label{xc10}
\frac{\partial p_{i/i-1}}{\partial t} = -\frac{\tau}{m}\frac{\partial
(F_ip_{i/i-1})}{\partial x_i}+\frac{1}{2m\eta}\frac{\partial^2 p_{i/i-1}}{\partial
x_i^2}.
\end{eqnarray}
This is the Fokker-Planck equation, where $\tau$ is the mean free time supposed to
be the time interval $t$ of the particle on each segment of its path. In view of the
Eq.(\ref{acc9x}), it is easy to show that this equation is also satisfied by
$p_k(x|a)$ if $x_i$ is replaced by $x$, the final position. Now let $n_a$ and $n_b$
be the particle density at $a$ and $b$, respectively. The following relationship
holds
\begin{eqnarray}                                            \label{xxc10x}
n_b=\sum_{k}n_ap_k(x|a)
\end{eqnarray}
which is valid for any $n_a$. This means ($n=n_b$ and $x=x_b$):
\begin{eqnarray}                                            \label{xc10x}
\frac{\partial n}{\partial t} = -\frac{\tau}{m}\frac{\partial (Fn)}{\partial
x}+\frac{1}{2m\eta}\frac{\partial^2 n}{\partial x^2},
\end{eqnarray}
which describes the time evolution of the particle density.

\section{Fick's laws of diffusion}
If the external force $F$ is zero, we get
\begin{eqnarray}                                            \label{c10xx}
\frac{\partial n}{\partial t} = \frac{1}{2m\eta}\frac{\partial^2 n}{\partial x^2}
\end{eqnarray}
This is the second Fick's law of diffusion. The first Fick's law Eq.(\ref{1}) can be
easily derived if we consider matter conservation $\frac{\partial n(x,t)}{\partial
t}= -\frac{\partial J(x,t)}{\partial x}$ where $J(x,t)$ is the flux of the particle
flow.

\section{Ohm's law of charge conduction}
Considering the charge conservation $\frac{\partial \rho(x,t)}{\partial t}=
-\frac{\partial j(x,t)}{\partial x}$, where $\rho(x,t)=qn(x,t)$ is the charge
density, $j(x,t)=qJ(x,t)$ is the flux of electrical currant and $q$ is the charge of
the currant carriers, we have, from Eq.(\ref{xc10x}),
\begin{eqnarray}                                            \label{xc10xxx}
j=\frac{\tau}{m}F\rho - \frac{1}{2m\eta}\frac{\partial \rho}{\partial x}
\end{eqnarray}
where $F=qE$ is the electrostatic force on the carriers and $E$ is the electric
field. If the carrier density is uniform everywhere, i.e., $\frac{\partial
n}{\partial x}=0$, we get the Ohm's law
\begin{eqnarray}                                            \label{xxc10xxx}
j=\frac{\tau}{m}qE\rho=\sigma E,
\end{eqnarray}
where $\sigma=\frac{q\rho\tau}{m}=\frac{nq^2\tau}{m}$ is the formula of electrical
conductivity widely used for metals.

\section{Passage from stochastic to regular dynamics}
The present probabilistic representation of dynamics using least action distribution
underlies a biased formalism of classical mechanics defined with the deterministic
Hamiltonian functional $H$. It is clear that Eqs.(\ref{wc6}) and (\ref{w6x}) do not
exist for the paths other than the unperturbed geodesics (least action ones). We have
in general $\delta A_{ab}(k) \neq 0$ which is either positive or negative, i.e.,
\begin{eqnarray}                                            \label{7aa}
\delta A_{ab}(k) &=& \int_a^b \left[\frac{\partial}{\partial t}\frac{\partial
L_{k}(t)}{\partial \dot{x}}-\frac{\partial L_{k}(t)}{\partial x}\right]\varepsilon
dt\neq 0
\end{eqnarray}
where $\varepsilon$ is an arbitrary variation of $x$ ($\varepsilon$ is zero at $a$
and $b$). Using the same mathematics as in \cite{Arnold}, we can prove that, if
$dA_{ab}(k)=\frac{\partial A_{ab}(k)}{\partial x}dx -\frac{\partial
A_{ab}(k)}{\partial \dot{x}}d\dot{x}>\;(or <)\;0$, a deformed Euler-Lagrange
equation follows
\begin{eqnarray}                                            \label{7ab}
\frac{\partial}{\partial t}\frac{\partial L_{k}(t)}{\partial \dot{x}}-\frac{\partial
L_{k}(t)}{\partial x}> 0\;\;(or <0).
\end{eqnarray}
Then with the help of the Legendre transformation $H=P\dot{x}-L_{k}$\cite{Arnold},
we can derive
\begin{eqnarray}                                            \label{1a}
\dot{x}=\frac{\partial H}{\partial P} \;\;and \;\; \dot{P}> \;(or <)\;
-\frac{\partial H}{\partial x},
\end{eqnarray}
which are the Hamiltonian equations for irregular dynamic systems along the paths with
non stationary action. The second equation above violates the Newton's second law.

However, as mentioned above, the average action has a stationary associated with
stationary path information as shown in Eq.(\ref{ws6x}). This ``least average
action'' can be used in the following calculations to derive an averaged version of
the equations of motion of classical mechanics.

\begin{eqnarray}                                            \label{7}
\delta A_{ab} &=& \sum p_k(b|a)\delta A_{ab}(k)+\sum \delta p_k(b|a)A_{ab}(k) \\
&=&\int_a^b \left[\left\langle\frac{\partial}{\partial t}\frac{\partial
L_{k}(t)}{\partial \dot{x}}\right\rangle-\left\langle\frac{\partial
L_{k}(t)}{\partial x}\right\rangle\right]\varepsilon dt+\frac{\delta H_{ab}}{\eta}.
\end{eqnarray}
Now taking into account Eq.(\ref{ws6x}), and using the usual
mathematics\cite{Arnold}, we obtain an extended version of the Euler-Lagrange
equations :
\begin{eqnarray}                                            \label{wc6x}
\left\langle\frac{\partial}{\partial t}\frac{\partial L_{k_{ab}}(t)}{\partial
\dot{x}}\right\rangle-\left\langle\frac{\partial L_{k_{ab}}(t)}{\partial
x}\right\rangle=0
\end{eqnarray}
and of the Hamiltonian equations
\begin{eqnarray}                                            \label{w6xx}
\langle\dot{x}\rangle=\left\langle\frac{\partial H}{\partial P}\right\rangle
\;\;,\;\; \langle\dot{P}\rangle=-\left\langle\frac{\partial H}{\partial
x}\right\rangle
\end{eqnarray}
where the average $\left\langle\ \right\rangle$ is taken over all the possible path
between two points.

At first glance, it may seems surprising that the non least action paths have non
zero probability of occurring and the fundamental equations such as Eqs.(\ref{wc6})
and (\ref{w6x}) be violated within classical mechanics. It should be noticed that
the probabilistic approach of this work addresses the {\it Hamiltonian systems
subject to thermal noise and chaotic motion}. So strictly speaking, the Hamiltonian
of the systems does not exist. It exists only in an average manner as shown by
Eq.(\ref{wc6x}) and Eq.(\ref{w6xx}). The random perturbation is of course
responsible for the violation of the fundamental equations of classical mechanics.
In fact, we can introduce a random force $R$ into the classical equations. For a
path whose action is not at stationary, Eq.(\ref{w6xx}) implies a Langevin equation:
\begin{eqnarray}                                            \label{wcx6x}
\dot{P}=-\frac{\partial H}{\partial x}+R
\end{eqnarray}
where $R$ is positive if $\delta A_{ab}(k)>0$ and negative if $\delta A_{ab}(k)<0$.
From Eq.(\ref{w6xx}), we must have $\left\langle R\right\rangle=0$.

Logically, this probabilistic dynamics should recover the regular dynamics when the
uncertainty tends to zero. In fact, the dispersion of action or the width of the
least action distribution can be measured by the variance
$\sigma^2=\overline{A^2}-\overline{A}^2=\overline{A^2}-A_{ab}^2$. From
Eqs.(\ref{c1x}), (\ref{c1xxx}), (\ref{c6x}) and (\ref{cx6x}), we get
\begin{eqnarray}                                            \label{w12}
\sigma^2=-\frac{\partial A_{ab}}{\partial \eta},
\end{eqnarray}

\begin{eqnarray}                                            \label{w7x}
A_{ab}=-\frac{\partial}{\partial\eta}\ln Z,
\end{eqnarray}
and
\begin{eqnarray}                                            \label{wc7}
H(a,b)=\ln Z+\eta A_{ab}=\ln Z-\eta \frac{\partial}{\partial\eta}\ln Z
\end{eqnarray}
When the system becomes less and less irregular, $\sigma^2$ diminishes and the paths
become closer and closer to the least action ones having stationary action
$A_{ab}^{stat}$. When $\sigma^2\rightarrow 0$, the significant contribution to the
partition function $Z$ comes from the geodesics having $A_{ab}^{stat}$, i.e.,
$Z\rightarrow exp[-\eta A_{ab}^{stat}]$. From Eq.(\ref{w7x}), $A_{ab}\rightarrow
A_{ab}^{stat}$. Then considering Eq.(\ref{wc7}), it is clear that the path
information $H(a,b)\rightarrow 0$. In this case, the stationary average action
Eq.(\ref{ws6x}) becomes $\delta A_{ab}^{stat}=0$, the usual action principle, and
Eqs.(\ref{wc6x}) and (\ref{w6xx}) will recover Eqs.(\ref{wc6}) and (\ref{w6x}). At
the same time, diffusion phenomena completely vanish and the diffusion laws are
replaced by the laws of regular mechanics.

\section{Concluding remarks}
We have presented an attempt to investigate diffusion phenomena with a method which
consists in placing hamiltonian systems under the random perturbation of thermal
noise and chaotic instability and addressing the irregular dynamics produced in this
way with a information approach. A path information is used to measure the dynamic
uncertainty associated with different possible transport paths in phase space
between two points. On the basis of the assumption that the paths can be physically
characterized by their action (defined by using the unperturbed Hamiltonian), we
show that the maximum path information yields a path probability distribution in
exponential of action.\footnote{We would like to note here that this exponential
transition probability distribution of action has nothing to do with the Feynman's
postulated quantum propagator proportional to the path integrals of an exponential
of action multiplied by the imaginary number $i$\cite{Feynman}. Although the
mathematics of this work is similar to that of the path integrals, the physics is
totally different. In path integral quantum theory, the transition probability
between two points is the absolute square of the propagator, hence it is not
necessarily exponential of action.} This least action distribution provides a simple
and general derivation of the Fokker-Planck equation, the Fick's laws, and the Ohm's
law. The mathematics of this derivation is not new. It can be found in many text
books treating diffusion on the basis of Brownian motion and Markovian process. The
new point of this work is using the information method combined with the action
principle underlying a transition probability in exponential of action and a
probabilistic formalism of dynamics for the hamiltonian systems under random
perturbation. The reasoning is simple and the mathematics is straightforward. The
usual assumptions\cite{Kubo,Zaslavsky} for the derivation of diffusion laws, e.g.,
Brownian motion, Markovian process, Gaussian random forces, or Kolmogorov
conditions, are unnecessary in this approach.

Although in this approach the action of individual paths is not at stationary, the
stationary path information assures that the action of the most probable paths has a
stationary and that the average action over all possible paths have also a
stationary under the constraint associated with dynamic uncertainty. This result
suggests the following analogy: as the Newton's law is the consequence of action
principle of regular hamiltonian systems, the diffusion laws can also be considered
as the consequences of this same principle which presents itself through the
stationary average action associated with irregular dynamic uncertainty.

The validity of the present work is of course limited by the mathematical tools we
use in this work. As far as we can see, the results of this work probably does not
hold if, 1) the dynamic uncertainty cannot be measured by Shannon information; 2)
the paths of the diffusants is not sufficiently smooth for the application of the
Euler method; 3) the paths do not allow continuous variational method. The latter
two cases may occur when, for example, the phase space is porously occupied such as
in fractal phase space.


\begin{thebibliography}{99}
\bibitem {Berthalot}
C.L. Berthalot, Eassai de Statique Chimique, Paris, France, 1803

\bibitem {Fick}
A. Fick, {\em Ann. Physik, Leipzig,} {\bf 170}(1855)59

\bibitem {Lebowitz}
J.L. Lebowitz and H. Spohn, Microscopic basis for Fick's laws of self-diffusion,
{\em J. Stat. Phys.,} {\bf 28}(1982)539-556

\bibitem {Bonetto}
F. Bonetto, J.L. Lebowitz, L. Rey-Bellet, Fourier's Law: a challenge for theorists,
In A. Fokas, A. Grigoryan, T. Kibble, and B. Zegarlinski (eds.), Mathematical
Physics 2000, pp. 128–150, London, 2000. Imperial College Press; math-ph/0002052

\bibitem {Bonetto2}
F. Bonetto, J.L. Lebowitz, J. Lukkarinen, Fourier's Law for a harmonic crystal with
selfd-consistent stochastic reservoirs, math-ph/0307035

\bibitem {Ohm}
N. Chernov, G. Eyink, J.L. Lebowitz, and Ya.G. Sinai, Derivation of Ohm's Law in a
Determinisitic Mechanical Model, {\em Phys. Rev. Lett.,} {\bf 70}(1993)2209

\bibitem {Kubo}
R. Kubo, M. Toda, N. Hashitsume, {\em Statistical physics II, Nonequilibrium
statistical mechanics,\/} Springer, Berlin, 1995

\bibitem {Zaslavsky}
G.M. Zaslavsky, Chaos, fractional kinetics and anomalous transport, {\em Physics
Reports,} {\bf 371}(2002)461

\bibitem {Wang04x}
Q.A. Wang, Maximum path information and the principle of least action for chaotic
system, {\em Chaos, Solitons $\&$ Fractals,} (2004), in press; cond-mat/0405373 and
ccsd-00001549

\bibitem {Dorfman}
J.R. Dorfmann, {\em An introduction to Chaos in nonequilibrium statistical
mechanics}, Cambridge University Press, 1999

\bibitem {Wang04xx}
Q.A. Wang, Action principle and Jaynes' guess method, cond-mat/0407515

\bibitem {Arnold}
V.L. Arnold, {\em Mathematical methods of classical mechanics,\/} Springer-Verlag,
New York, 1989

\bibitem {Feynman}
R.P. Feynman, {\em Quantum mechanics and path integrals,\/} McGraw-Hill Publishing
Company, New York, 1965


\end{thebibliography}
\end{document}